\documentstyle[12pt]{article}
\newcommand{\expect}[1]{\langle{#1}\rangle}
\newcommand{\dm}{\partial_{\mu}}
\newcommand{\vn}{\vec{n}}
\newcommand{\inv}[1]{{1 \over {#1}}}
\newcommand{\half}{{1\over 2}}

\newcommand{\qint}{\int{d^{3}q \over (2\pi)^3}}
\newcommand{\SR}{S^{1}\times{\bf R}^{2}}
\newcommand{\RS}{S^{2}\times{\bf R}^{1}}
\newcommand{\ssl}{\mbox{[}}
\newcommand{\ssr}{\mbox{]}}

\newcommand{\NN}{\nonumber}

\newcommand{\quadfour}{\quad\quad\quad\quad}

\begin{document}
\title{
\begin{flushright}
{\small YITP-K-1082}
\end{flushright}
\vspace{1cm}
Finite-size Analysis of $O(N)$ Nonlinear $\sigma$
Model on Semi-compact Spaces}
\author{Akira FUJII
\thanks{e-mail address: {\tt fujii@jpnyitp.bitnet}}\\
{\sl Yukawa Institute for Theoretical Physics,} \\
{\sl Kyoto University,} \\
{\sl Kyoto 606-01, Japan.}}
\date{August 3, 1994}
\maketitle
\begin{abstract}
Fisher's phenomenological renormalization method is used to
calculate the mass gap and the correlation length of the $O(N)$
nonlinear $\sigma$ model on a semi-compact space $\SR$. This shows
that the ultraviolet momentum cut-off does not conflict with the
infrared cut-off along the $S^{1}$ direction. The mass gap on
$\RS$ is also discussed.
\end{abstract}
\newpage
\baselineskip=24pt
The $O(N)$ nonlinear $\sigma$ (NL$\sigma$) model on a semi-compact space
$\SR$ has been intensively studied recently in the context of
two-dimensional quantum spin systems\cite{spin} and in researching the
critical phenomena in three dimensions\cite{fujii}\cite{rajeev}.
The calculation of the partition functions and correlation length (or
mass gap) is done with the ultraviolet (UV) momentum
cut-off in the ${\bf R}^{2}$ direction and usually without any UV momentum
cut-off along the $S^{1}$ direction being
imposed\cite{spin}\cite{sachdev}.
This procedure is quite natural in the view of the imaginary time
path integral method, which discritizes only the spatial coordinates
 and leaves the temporal one continuous.
However, this scheme might conflict with the infrared (IR)
cut-off along the $S^{1}$ direction because
of its unrenormalizability.

In this paper, we consider the gap
equation on the lattice and make use of Fisher's phenomenological
renormalization (PR) method\cite{bresin}\cite{fisher}
to clarify the above point. The Pr method is a renormalization
method in the real space based on a hypothesis of the pseudo-scale
invariance for large but finite systems. In this scheme, we have the
advantage that the UV cut-off does not appear explicitly.
 In terms of the PR method, the correlation length can be calculated
exactly on $S^{1}\times S^{1}\times{\bf R}^{1}$ or $S^{1}\times
S^{1}\times S^{1}$\cite{bresin}.
We apply the PR method to the $O(N)$ NL$\sigma$ model
on $\SR$ and $\RS$. These PR procedures reproduce
the mass-gap obtained previously
in the cut-off regularization\cite{spin}.
Therefore, we consider that the UV cut-off regularization
does not conflict with the IR cut-off along the $S^{1}$ or $S^{2}$
direction. Of course, it is widely believed that the physical
quantities
can be calculated independently of the regularization
in renormalizable systems. Our results are consistent with this
intuition despite its unrenormalizablilty in the ordinary meaning.

We consider the $O(N)$ NL$\sigma$ model on a semi-compact space
$\SR$ with the radius of $S^{1}$, $L$.
The partition function is given by
\begin{equation}
Z=\int D\vn(x)
\exp\left(-{N\over 2g}\int_{\SR}d^{3}x(\dm \vn(x))^{2}\right),
\end{equation}
where $\vn(x)$ is an $N$-dimensional vector
field normalized to $(\vn(x))^{2}=1$.
This model is believed to describe the long range behavior of the
two-dimensional antiferromagnetic Heisenberg model as
\begin{equation}
H=J\sum_{\langle ij \rangle}\vec{S}_{i}\cdot\vec{S}_{j},
\end{equation}
with large spin $S$ at finite temperature.
We now follow the PR method. To begin
with we solve the constraint $(\vn(x))^{2}=1$ by means of the
auxiliary field as
\begin{eqnarray}
Z&=&\int D\vn(x)D\mu(x)\exp\left(-{N\over 2g}\int_{\SR}d^{3}x
\ssl(\dm\vn(x))^{2}
+\mu(x)((\vn(x))^{2}-1)\ssr\right), \NN \\
&=&\int D\mu(x)\exp(-(N/2)S_{\it eff}), \label{effective}
\end{eqnarray}
where we denote
\begin{equation}
S_{\it eff}=-\inv{g}\int_{\SR}d^{3}x\mu(x)+\log\det(-\partial^{2}+\mu(x)).
\end{equation}
We consider the large $N$ limit, which enables us to make use of the
saddle point method\cite{arefeva}.
If we impose the periodic boundary condition in the direction of
$S^{1}$, the gap equation on $\SR$ with lattice regularization is given by
\begin{eqnarray}
\inv{g}&=&\qint \inv{\xi_{L}^{-2}+2\sum_{\mu=0}^{2}(1-\cos q_{\mu})}
\sum_{n\in {\bf Z}}(2\pi)\delta\left(q_{0}-{2\pi\over L}n\right) \NN \\
&=&\qint\inv{\xi_{L}^{-2}+2\sum_{\mu=0}^{2}(1-\cos q_{\mu})}
\sum_{n\in{\bf Z}}e^{iq_{0}nL}, \label{anten}
\end{eqnarray}
where we apply Poisson's summation formula and
$\xi_{L}=(\expect{\mu(x)})^{-1/2}$ is denoted as
the correlation length in $\SR$, which is a function of $L$ and $g$.
We put $\xi_{\infty}=\lim_{L\rightarrow\infty}\xi_{L}$, which
is divergent when the
coupling constant $g$ corresponds to the critical value $g_{c}$.
Here we denote
$g_{c}$ as the UV fixed point of the renormalization group.
If we subtract the equation (\ref{anten}) for $\xi_{\infty}$ from that for
$\xi_{L}$, the difference gives
\begin{eqnarray}
&& (\xi_{\infty}^{-2}-\xi_{L}^{-2})\qint
\inv{\xi_{\infty}^{-2}+2\sum_{\mu=0}^{2}(1-\cos q_{\mu})}
\inv{\xi_{L}^{-2}+2\sum_{\mu=0}^{2}(1-\cos q_{\mu})}\NN \\
&&\quadfour +\qint\sum_{n\neq0}
{e^{iq_{0}nL}\over\xi_{L}^{-2}+2\sum_{\mu=0}^{2}(1-\cos q_{\mu})}=0.
\end{eqnarray}
If we choose the coupling constant $g$ to be close to the critical value
$g_{c}$, we can replace the propagator $1/(\xi_{L}^{-2}+\sum(1-\cos q_{\mu}))$
by $1/(\xi_{L}^{-2}+q^{2})$ and expand the momentum region $\ssl-\pi,\pi\ssr$
to $(-\infty,\infty)$.
Therefore the  first term can be further simplified as
\begin{equation}
(\xi_{\infty}^{-2}-\xi_{L}^{-2})
\qint\inv{\xi_{\infty}^{-2}+q^{2}}\inv{\xi_{L}^{-2}+q^{2}}=
(\xi_{\infty}^{-2}-\xi_{L}^{-2}){\Gamma(1/2)\over (4\pi)^{3/2}}
\int^{1}_{0}{dt\over(t\xi_{L}^{-2}+(1-t)\xi_{\infty}^{-2})^{1/2}}.
\end{equation}
The second term  can be also simplified as
\begin{equation}
\sum_{n\neq 0}\qint {e^{iq_{0}nL}\over\xi_{L}^{-2}+q^2}=
\inv{L}\int^{\infty}_{0}dte^{-t(L/\xi_{L})^{2}}u(t),
\end{equation}
where we put $u(t)=\sum_{n\neq 0}e^{-n^2/4t}/(4\pi t)^{3/2}$.
Therefore, the gap equation is reduced to
\begin{eqnarray}
&&\left[ \left({L\over\xi_{\infty}}\right)^{2}-
\left({L\over\xi_{L}}\right)^{2}\right]
\left({\xi_{L}\over L}\right)
{\Gamma(1/2)\over (4\pi)^{3/2}}
\int^{1}_{0}dt\inv{\sqrt{(t(\xi_{L}/\xi_{\infty})+(1-t))}}
\NN \\ && \quadfour
+\int^{\infty}_{0}dte^{-t(L/\xi_{L})^{2}}u(t)=0.
\end{eqnarray}
If we choose the coupling constant $g$ to be the critical value
$g_{c}$, the correlation length in ${\bf R}^{3}$ becomes infinite and
the gap equation simplifies to
\begin{equation}
\sum_{n=1}^{\infty}{(\xi_{L}/L)\over n}e^{-n/(\xi_{L}/L)}=
-{\xi_{L}\over L}\log (1-e^{-L/\xi_{L}})=
\half,
\end{equation}
which reproduces the mass gap obtained in Ref.\cite{spin} as
\begin{equation}
\xi_{L}/L=1/(2\log{\sqrt{5}+1\over 2}).
\end{equation}
{}From this mass gap, the finite size correction can be calculated as
\begin{equation}
F_{L}-F_{\infty}={4N\over 5}{\zeta(3)\over L^{3}},
\end{equation}
making use of the addition formula of Roger's polylogarithmic
function\cite{sachdev}.

We can also calculate the correlation length in
this PR procedure as follows.

The UV stable critical coupling constant $g_{c}$
in ${\bf R}^{3}$ is given by
\begin{equation}
\inv{g_{c}}=\int_{{\bf R}^{3}}{d^{3}q\over (2\pi)^{3}}\inv{q^{2}}. \label{gc}
\end{equation}
If we subtract Eq. (\ref{gc}) from Eq. (\ref{anten}),
we obtain the equation
\begin{eqnarray}
\inv{g}-\inv{g_{c}}&=&-\inv{\xi_{L}^{2}}
\qint\inv{q^{2}}\inv{\xi_{L}^{-2}+q^{2}}+
2\sum_{n=1}^{\infty}\qint{e^{iq_{0}nL}\over \xi_{L}^{-2}+q^{2}}, \NN \\
&=&-\inv{4\pi\xi_{L}}-\inv{2\pi L}\log (1-e^{-L/\xi_{L}}), \label{corr}
\end{eqnarray}
which determines the correlation length $\xi_{L}$ on $\SR$.
We solve Eq.(\ref{corr}) in two regions of the coupling constant
$g$ in the large $L$ ({\it i.e.} low temperature) limit.
First, we consider the region $g\ll g_{c}$, which is called the {\it
renormalized classical region} in Ref.\cite{spin}.
In this region,
the correlation length is given by
\begin{equation}
\xi_{L}=L\exp\left(2\pi L\left(\inv{g}-\inv{g_{c}}\right)\right).
\end{equation}

Second we consider the region $g\gg g_{c}$, which is called the {\it
quantum disordered region} in Ref.\cite{spin}. In this region, the
solution of Eq. (\ref{corr}) is
\begin{equation}
\xi_{L}^{-1}={2\pi\over g_{c}}\left(1-{g_{c}\over g}\right),
\end{equation}
which is independent of the size of $S^{1}$, $L$.
These expressions for the correlation lengths in the two regions correspond
to those obtained by the momentum cut-off method\cite{spin}.

{}From the above calculations, we can see that the UV momentum cut-off
procedure does not conflict with the IR cut-off $L$ in the case of
the $O(N)$ NL$\sigma$ model on $S^{1} \times {\bf R}^{2}$.

Furthermore, we can discuss the dependence of the correlation length in
$S^{2} \times {\bf R}$, with
$R$, which is the radius of $S^{2}$. In this
case, denoting the correlation length with the radius $R$
as $\xi_{R}$, the gap equation is given by
\begin{eqnarray}
\inv{g}&=&\inv{4\pi R^{2}}\sum_{n=0}^{\infty}
{2n+1\over \xi_{R}^{-2}+p^{2}+n(n+1)/R^{2}} \NN \\
&=&\int{d^{3}\vec{p}\over (2\pi)^{3}}\inv{(\xi_{R}^{-2}-\inv{4R^{2}})
+\vec{p}^{2}}
+2\sum_{n=1}^{\infty}(-)^{n}\int{d^{3}\vec{p}\over (2\pi)^{3}}
\inv{(\xi_{R}^{-2}-\inv{4R^{2}})+\vec{p}^{2}}
\cos(2\pi Rn|{\bf p}_{\perp}|). \NN \\
&& \label{s2r1}
\end{eqnarray}
Here we make a use of Poisson's summation formula again and denote the
three-dimensional vector $\vec{p}=({\bf p}_{\perp},p)$.
In a similar manner to the case of $\SR$, we subtract Eq.
(\ref{s2r1}) for $R=\infty$ from that for finite $R$. If we put the
coupling constant $g$ at the critical value $g_{c}$, the difference
gives
\begin{equation}
-{m\over 4\pi}+2\sum_{n=1}^{\infty}\int {d^{3}\vec{p}\over (2\pi)^{3}}
\inv{{\vec p}^{2}+m^{2}}(-)^{n}\cos(2\pi Rn|{\bf p}_{\perp}|)=0, \label{r2s1c}
\end{equation}
where we put $m^{2}=\xi_{R}^{-2}-(1/4R^{2})$. The solution of
Eq.(\ref{r2s1c}) is \cite{fujii}
\begin{equation}
m=0 \quad\quad\quad i.e. \quad \xi_{R}=2R, \label{gaps2r1}
\end{equation}
which is consistent with the expected value of the conformal
coupling in the three-dimensional space\cite{cardy}.
The conformally invariant free scalar theory in a three-dimensional
space ${\it M}$ is given by
\begin{equation}
S=\half\int_{\it M}\ssl
g^{\mu\nu}\partial_{\mu}\phi\partial_{\nu}\phi+
\eta R^{(2)}\phi^{2}
\ssr \sqrt{g}d^{3}x,
\end{equation}
where $\eta=1/8$ and $R^{(2)}$ is the scalar curvature of $M$.
 If $M=S^{2} \times {\bf R}$, the conformal coupling takes the value of $\eta
R^{(2)}=1/2R$,
which corresponds to the obtained value of $\xi_{R}^{-1}$ in
Eq. (\ref{gaps2r1}).
Therefore with the
coupling constant $g_{c}$, the $O(N)$ NL$\sigma$ model is conformally
invariant on $S^{2} \times {\bf R}$.

Even though we do not have the exact form of the correlation length on
$S^{2}\times S^{1}$, even
with the critical coupling constant $g_{c}$, we expect that
the UV cut-off regularization method does not conflict with the
IR cut-off in this case either, and therefore that
the UV cut-off regularization
is available for $S^{2} \times S^{1}$.

{\bf Acknowledgment}%

The author would like to acknowledge Professors Hikaru Kawai and
Hisashi Yamamoto for useful comments. He also thanks Dr. Michael Gagen
for reading the manuscript.

\end{document}